\documentclass[12pt]{article}
  \usepackage{amsfonts}
  \usepackage{amsmath}
\usepackage{amssymb}
\usepackage{amscd}
\usepackage{graphicx}
\begin{document}

~~
\bigskip
\bigskip
\begin{center}
{\Large {\bf{{{Twist deformations of Newtonian Schwarzschild-(Anti-)de Sitter classical system}}}}}
\end{center}
\bigskip
\bigskip
\bigskip
\begin{center}
{{\large ${\rm {Marcin\;Daszkiewicz}}$}}
\end{center}
\bigskip
\begin{center}
\bigskip

{ ${\rm{Institute\; of\; Theoretical\; Physics}}$}

{ ${\rm{ University\; of\; Wroclaw\; pl.\; Maxa\; Borna\; 9,\;
50-206\; Wroclaw,\; Poland}}$}

{ ${\rm{ e-mail:\; marcin@ift.uni.wroc.pl}}$}

\end{center}
\bigskip
\bigskip
\bigskip
\bigskip
\bigskip
\bigskip
\bigskip
\bigskip
\bigskip
\begin{abstract}
In this article we provide three new twist-deformed Newtonian Schwarzschild-(Anti-)de Sitter models. They are defined on the Lie-algebraically as well as on the canonically noncommutative space-times respectively. Particularly we find the corresponding Hamiltonian functions and the proper equations of motion. The relations between the models are discussed as well. 
\end{abstract}
\bigskip
\bigskip
\bigskip
\bigskip
\eject

\section{{{Introduction}}}
$$~~~$$

The Schwarzschild-(Anti-)de Sitter metric plays an important role for the most general solution of the vacuum Einstein equation with point-like massive source and nonvanishing cosmological constant $\Lambda$. It has been proposed many years ago by Kottler \cite{kottler} but despite of that it still seems to be quite interesting. 
Since the SdS\footnote{The hashes SdS and S(A)dS mean Schwarzschild-de Sitter and  Schwarzschild-(Anti-)de Sitter metrics respectively.} tensor contains the $\Lambda$-terms it can describe the space-time geometry near the heavy object in the Universe of which expansion is 
generated by cosmological repulsive force. Recently, for example (see paper \cite{bending}) there has been studied the impact of such a force on the light bending. Besides the classical tests\footnote{It has been studied the effects of cosmological constant $\Lambda$ on Mercury’s perihelion precession and light bending in the context of the Newtonian limit of SdS space-time.} of the Newtonian Schwarzschild-de Sitter space have been performed in article \cite{tests}.

Regardless of the above considerations there appeared a lot of papers concerning the influence of space-time noncommutativity on the dynamics of physical systems. The proper investigation has been accomplished in the theoretical field (see e.g. \cite{prefield}-\cite{field}), chaos modeling (see e.g. \cite{def1}-\cite{defn}) as well as in the classical and quantum mechanical (see e.g. \cite{mech}-\cite{qm}) context. Consequently, it seems to be quite vital to study the ascendancy of quantum space on the structure of the relativistic and nonrelativistic Schwarzschild-(Anti-)de Sitter systems as well. It should be noted that such a research has been achieved at both velocity scale levels only in the case of canonical\footnote{In accordance with the Hopf-algebraic classification of all
deformations of relativistic \cite{class1} and nonrelativistic
\cite{class2} symmetries, one can distinguish three basic types
of space-time noncommutativity (see also \cite{nnh} for details): canonical \cite{oeckl}-\cite{dasz1}, Lie-algebraic \cite{dasz1}-\cite{lie1}
and Quadratic deformation of Minkowski and Galilei  spaces \cite{dasz1}, \cite{lie1}-\cite{paolo}.} deformation in articles \cite{bad2}, \cite{badn}. 

In this paper, we provide the three noncommutative Newtonian SdS and S(A)dS models. All of them are defined on the twisted\footnote{For details concerning the twist deformation of Hopf algebras see \cite{drin} and \cite{qg}.} Galilei space-times such as canonically as well as Lie-algebraically deformed spaces. Particularly, we provide the proper Hamiltonian functions and the corresponding equations of motion. Apart of that we dynamically couple the constructed models each other with use of the so-called active control synchronization procedure \cite{conect1}, \cite{conectn}\footnote{By synchronization we mean a dynamical coupling of particles moving in the presence of different (deformed) dynamics such that their phase space trajectories for large times of the evolution become the same.}. In such a way we introduce and analyze the nonrelativistic systems which formally describe the impact of both transplanckian (noncommutativity) and cosmological (presence of parameter $\Lambda$) distance scales on dynamics of the Newtonian S(A)dS models. The especially interesting seems to be the Lie-algebraically noncommutative systems due to the fact, that they provide in natural way the deformation parameter $\kappa$ which plays a role of Planck mass \cite{planck}. It is commonly belived that studies  on such a type of space-time noncommutativity might shed some additional light on for example the properties of Quantum Gravity Theory \cite{Qgrav}. For this reason the proposed in this article models have a chance to give an alternative description of nonrelativistic Quantum Gravity effects in cosmological context \cite{QGC}. 

The paper is organized as follows. In second Section we recall the basic facts concerning the Schwarzschild-(Anti-)de Sitter metric and its nonrelativistic limit. Section three is devoted to the canonically and Lie-algebraically twist deformations of Galilei Hopf algebra. In Section four we provide the corresponding S(A)dS systems while the relations between models and their synchronizations are discussed in Section five. The final remarks close the paper.

\section{Newtonian Schwarzschild-(Anti-)de Sitter classical model}

Let us start with vacuum Einstein equation for nonvanishing cosmological constant $\Lambda$
\begin{equation}
R_{\mu\nu} - \frac{1}{2}g_{\mu\nu}R = \Lambda g_{\mu\nu}\;.\label{vacum}
\end{equation}
Its most general spherically symmetric, so-called Schwarzschild-de Sitter $(\Lambda>0)$ or Schwarzschild-Anti-de Sitter $(\Lambda<0)$ solution take the form
\begin{equation}
ds^2 = -f(r) dt^2 + \frac{dr^2}{f(r)} + r^2d\Omega\;, \label{solution}
\end{equation}
with
\begin{equation}
f(r) = 1 - \frac{2GM}{r} - \frac{\Lambda c^2}{3}\bar{r}^2\;,\label{fun}
\end{equation}
where symbols $G$ and $M$ denote the Newton constant and mass of point-like source respectively, while $\bar{r} = [\;x_1,x_2,x_3\;]$. Besides, one can check that the nonrelativistic limit of the above metric
generates the following potential 
\begin{equation}
\phi_{SdS/AdS} = -\frac{GM}{r} - \frac{\Lambda c^2}{3}\bar{r}^2\;,\label{potential}
\end{equation}
which leads to the proper Schawrzschild-(A)de Sitter Hamiltonian function\footnote{Symbol $m$ denotes the mass of probing particle.}
\begin{equation}
H_{SdS/AdS} = \frac{\bar{p}^2}{2m} -\frac{GmM}{r} - \frac{m\Lambda c^2}{3}\bar{r}^2\;,\label{ham}
\end{equation}
as well as to the canonical equations of motion given by
\begin{equation}
\left\{\begin{array}{rcl}
\dot{p}_1 &=& -G\frac{mM}{r^3}x_1 + \frac{2}{3}m\Lambda c^2 x_1 \;\\[5pt]
&~~&~\cr
\dot{p}_2 &=& -G\frac{mM}{r^3}x_2 + \frac{2}{3}m\Lambda c^2 x_2 \;\\[5pt]
&~~&~\cr
\dot{p}_3 &=& -G\frac{mM}{r^3}x_3 + \frac{2}{3}m\Lambda c^2 x_3 \;\\[5pt]
&~~&~\cr
\dot{x}_1 &=& \frac{p_1}{m} \;\\[5pt]
&~~&~\cr
\dot{x}_2 &=& \frac{p_2}{m} \;,\\[5pt]
&~~&~\cr
\dot{x}_3 &=& \frac{p_3}{m} \;.
\label{equations1}\end{array}\right.
\end{equation}
Of course, for $M=0$ and $\Lambda \neq 0$ we get from the above model the attractive or repulsive oscillator system, while for $M \neq 0$ and $\Lambda=0$ we reproduce the Newtonian model of particle moving in the central gravitational field.

\section{Twist deformations of Galilei Hopf algebra}

In accordance with the twist procedure \cite{drin}, the algebraic sector of deformed Hopf structure remains classical. However, the corresponding coproducts transform in nontrivial way as follows\footnote{$\kappa$ denotes the deformation parameter.}
\begin{equation}
\Delta _{(0)}(a) \to \Delta _{(\kappa)}(a) = \mathcal{F}_{\kappa }\cdot
\,\Delta _{(0)}(a)\,\cdot \mathcal{F}_{\kappa }^{-1}\;\;\;;\;\;\;\Delta _{(0)}(a) = a\otimes 1 + 1\otimes 1\;,\label{transform}
\end{equation}
with twist factor
\begin{equation}
\mathcal{F}_{\kappa }=\exp i r_\kappa\;,  \label{factor}
\end{equation}
where $r_\kappa \in {\cal U}_0({\cal A})\otimes {\cal U}_0({\cal A})$ denotes so-called classical $r$-matrix satisfying the classical Yang-Baxter equation (CYBE) of the form
\begin{equation}
[[\;r_{\kappa},r_{\kappa}\;]] = 0\;;
\label{cybe}
\end{equation}
the symbol $[[\;\cdot\;,\;\cdot\;]]$ plays a role of so-called Schouten bracket \cite{qg}.

\subsection{Canonical twist deformation of Galilei Hopf algebra}

The canonically deformed Galilei Hopf algebra ${\cal U}_{\theta_{ij}}({\cal G})$ has been provided in article \cite{dasz1}
by the proper contraction of its relativistic counterpart. It is described by the classical $r$-matrix of the form\footnote{$a\wedge b = a\otimes b - b\otimes a.$}
\begin{eqnarray}
r_{\theta^{ij}} = \frac{1}{2}\theta^{ij}\Pi_i\wedge \Pi_j\;, \label{canr}
\end{eqnarray}
with $\Pi_i$ denoting the momentum generators. Then, in accordance with the general twist procedure it is given by the classical algebraic sector
\begin{eqnarray}
&&\left[\, K_{ij},K_{kl}\,\right] =i\left( \delta
_{il}\,K_{jk}-\delta
_{jl}\,K_{ik}+\delta _{jk}K_{il}-\delta _{ik}K_{jl}\right) \;,  \label{ff} \\
&~~&  \cr &&\left[\, K_{ij},V_{k}\,\right] =i\left( \delta
_{jk}\,V_i-\delta _{ik}\,V_j\right)\;\; \;, \;\;\;\left[
\,K_{ij},\Pi_{\rho }\right] =i\left( \eta _{j \rho }\,\Pi_{i }-\eta
_{i\rho }\,\Pi_{j }\right) \;, \label{nnn}
\\
&~~&  \cr &&\left[ \,V_i,V_j\,\right] = \left[ \,V_i,\Pi_{j
}\,\right] =0\;\;\;,\;\;\;\left[ \,V_i,\Pi_{0 }\,\right]
=-i\Pi_i\;\;\;,\;\;\;\left[ \,\Pi_{\rho },\Pi_{\sigma }\,\right] =
0\;,\label{ff1}
\end{eqnarray}
where $K_{ij}$, $\Pi_0$ and $V_i$ can be identified with rotation, time translation and boost operators as well as by the following
twisted coproducts
\begin{eqnarray}
&&\Delta_{(\theta^{ij})}(\Pi_\mu)=\Delta_{(0)}(\Pi_\mu)\;\;\;,\;\;\;
\Delta _{(\theta^{ij})}(V_i) =\Delta _{(0)}(V_i)\;, \label{dlww3v}\\
&~~~&  \cr \Delta _{(\theta^{ij})}(K_{ij})
&=&\Delta _{(0)}(K_{ij})-%
\theta ^{k l }[(\delta_{k i}\Pi_{j }-\delta_{k j
}\,\Pi_{i})\otimes \Pi_{l }\nonumber\\
&&\qquad\qquad\qquad\qquad\qquad+\Pi_{k}\otimes (\delta_{l
i}\Pi_{j}-\delta_{l j}\Pi_{i})]\;.\label{zadruzny}
\end{eqnarray}
Besides, it should be noted that the corresponding quantum space-time is given by
\begin{eqnarray}
\left[\;{t},{x}_i\;\right] = 0\;\;\;,\;\;\;\left[\;{x}_i,{x}_j\;\right] = i{\theta_{ij}}\;, \label{eq5}
\end{eqnarray}
and for deformation parameter $\theta_{ij}$ approaching zero it becomes classical.

\subsection{Lie-algebraic twist deformations of Galilei Hopf structure}

The two Lie-algebraically twist-deformed Galilei Hopf structures ${\cal U}_{\kappa_1}({\cal G})$ and ${\cal U}_{\kappa_2}({\cal G})$ have been
introduced in article \cite{dasz1} as well and they are described by the following $r$-matrices\footnote{The indexes $k$, $l$, $\gamma$ are fixed, spatial and different.}
\begin{equation}
r_{\kappa_1}=\frac{1}{2\kappa_3}\Pi_{\gamma}\wedge K_{kl} \;,
\label{r1}
\end{equation}
and\footnote{The indexes $k$, $l$ are fixed, spatial and different.}
\begin{equation}
r_{\kappa_2}=\frac{1}{2\kappa_3}\Pi_{k}\wedge V_{l} \;,
\label{r2}
\end{equation}
respectively. Their algebraic sectors remain classical (see formulas (\ref{ff})-(\ref{ff1})) while the coproducts are given by\footnote{$a\perp b = a\otimes b+b\otimes a$.},\footnote{$\psi_\lambda =\eta_{\nu
\lambda }\eta_{\beta \mu}-\eta_{\mu \lambda }\eta_{\beta
\nu}\;,\;\chi_\lambda =\eta_{\nu \lambda }\eta_{\alpha
\mu}-\eta_{\mu \lambda }\eta_{\alpha \nu}$.}
\begin{eqnarray}
 \Delta_{(\kappa_1)}(\Pi_0)&=&\Delta _{(0)}(\Pi_0)\;,\label{ffdlww2.2}\\
&~~&  \cr \Delta_{(\kappa_1)}(\Pi_i)&=&\Delta _{(0)}(\Pi_i)+\sin\left( \frac{1}{\kappa_1}
\Pi_\gamma \right)\wedge
\left(\delta_{k i}\Pi_l -\delta_{l i}\Pi_k \right)\label{america1}\\
&+&\left[\cos\left(\frac{1}{\kappa_1}  \Pi_\gamma \right)-1\right]\perp \left(\delta_{k i}\Pi_k
+\delta_{l i}\Pi_l \right)\;, \nonumber\\
~~ \Delta_{(\kappa_1)}(K_{ij})&=&\Delta_{(0)}(K_{ij})+K_{k l }\wedge \frac{1}{\kappa_1}
\left(\delta_{i
\gamma }\Pi_j-\delta_{j \gamma }\Pi_i\right)\nonumber\\
&+&i\left[K_{ij},K_{k l }\right]\wedge
\sin\left(\frac{1}{\kappa_1} \Pi_\gamma \right) \notag \\
&+&\left[\left[%
K_{ij},K_{k l }\right],K_{k l }\right]\perp
\left[\cos\left(\frac{1}{\kappa_1}  \Pi_\gamma  \right)-1\right]  \label{ffdlww2.22} \\
&+&K_{k l }\sin\left(\frac{1}{\kappa_1} \Pi_\gamma \right)\perp
\frac{1}{\kappa_1} \left(\psi_\gamma \Pi_k -\chi_\gamma \Pi_l \right) \notag \\
&+&\frac{1}{\kappa_1} \left(\psi_\gamma \Pi_l +\chi_\gamma \Pi_k \right)\wedge K_{k
l }\left[\cos\left(\frac{1}{\kappa_1} \Pi_\gamma \right)-1\right]\;,
  \notag\\
\Delta_{(\kappa_1)}(V_i)&=&\Delta_{(0)}(V_i) +i\left[V_i,K_{k l }\right]\wedge
\sin\left(\frac{1}{\kappa_1} \Pi_\gamma \right)
\label{sfifdlww2.22}\\
&+&\left[\left[
V_i,K_{k l }\right],K_{k l }\right]\perp \left[\cos\left(\frac{1}{\kappa_1} \Pi_\gamma
\right)-1\right]\;,  \nonumber
\end{eqnarray}
in case of the first quantum group, and
\begin{eqnarray}
~~~~~~~~~~~~\Delta_{({\kappa_2})}(\Pi_{0})&=&\Delta _{(0)}(\Pi_0) +
\frac{1}{{{\kappa_2}}} \Pi_l \wedge \Pi_k\;,\label{coppy1}\\
 &~~&  \cr
\Delta_{({\kappa_2})}(\Pi_i)&=&\Delta
_{(0)}(\Pi_i)\;\;\;,\;\;\;\Delta_{({\kappa_2})}(V_i)=\Delta_{(0)}(V_i)\;,\label{cop0}\\
 &~~&  \cr
\Delta_{({\kappa_2})}(K_{ij})&=&\Delta_{(0)}(K_{ij})+
\frac{i}{{{\kappa_2}}}\left[K_{ij},V_k\right]\wedge \Pi_l +
\frac{1}{{{\kappa_2}}}V_k \wedge(\delta_{il}\Pi_j
-\delta_{jl}\Pi_i)  \;, \label{coppy100}
\end{eqnarray}
for the second, ${\cal U}_{\kappa_2}({\cal G})$ Hopf structure. One can also check that that the corresponding quantum space-times look as follows
\begin{equation}
[\, x_{i },x_{j }\,] = \frac{i}{\kappa_1} \delta_{\gamma j}(
 \delta _{k i }x_{l }- \delta_{l i }x_{k }) +\frac{i}{\kappa_1} \delta_{\gamma
i}(\delta _{ l j }x_{k } - \delta _{k j }x_{l } )
\;\;\;,\;\;\;[\,t,x_i\,] =  0\;, \label{ssstar}
\end{equation}
and
\begin{equation}
[\, x_{i },x_{j }\,]=
\frac{i}{{\kappa_2}}t(\delta _{l i }\delta_{k j }-\delta _{
ki }\delta _{l j
})\;\;\;,\;\;\;[\,t,x_i\,] = 0 \;.
\label{ysesstar}
\end{equation}
Obviously, for deformation parameters $\kappa_1$ and $\kappa_2$ running to infinity the above relations become commutative.

\section{Twisted Newtonian Schwarzschild-(Anti-)de Sitter classical systems}
$~~$\\

\subsection{Canonical deformation}

In the first step of our construction we put in canonical commutation relations  (\ref{eq5}) the parameter $\theta_{12}$ equal to $\theta$ and $\theta_{23}=0=\theta_{13}$; in such a way we get\footnote{$\{\cdot,\cdot\}=\frac{1}{i}[\cdot,\cdot]$.}
\begin{eqnarray}
\{\;\hat{x}_{1 },\hat{x}_{2 }\;\} = \theta\;\;\;,\;\;\;\{\;\hat{x}_{1 },\hat{x}_{3 }\;\} =  0 = \{\;\hat{x}_{2 },\hat{x}_{3}\;\} \;. \label{newspace}
\end{eqnarray}
Next, we extend the above structure to the whole phase space as follows \cite{miao}\footnote{It is the most simple phase space for canonical space-time noncommutativity which satisfy the Jacobi condition.}
\begin{eqnarray}
\{\;\hat{x}_{1 },\hat{x}_{2 }\;\} = \theta\;\;\;,\;\;\;\{\;\hat{x}_{i },\hat{p}_{j}\;\} = \delta_{ij}\;\;\;,\;\;\;\{\;\hat{p}_{i},\hat{x}_{j}\;\} = 0 =
\{\;\hat{x}_{1 },\hat{x}_{3 }\;\} = \{\;\hat{x}_{2 },\hat{x}_{3}\;\}\;,\label{phasespace}
\end{eqnarray}
with $\hat{p}_i$ denoting the canonical momentum conjugated to ${\hat x}_i$-variable. By direct calculation one can check that the brackets (\ref{phasespace}) satisfy the Jacobi
identity and for deformation parameter $\theta$ approaching zero they reproduce the classical ones 
\begin{eqnarray}
\{\;{x}_{i },{p}_{j }\;\} = \delta_{ij}\;\;\;,\;\;\;\{\;{x}_{i },{x}_{j }\;\} = 0 = \{\;{p}_{i},{p}_{j}\;\}
\;. \label{classical}
\end{eqnarray}
Besides, it should be noted that the quantum variables $({\hat x}_i,{\hat p}_i)$ can be represented in terms of commutative ones $({x}_i,{p}_i)$ 
with use of Bopp shift \cite{bopp}
\begin{eqnarray}
\hat{x}_{1 } = x_1-\frac{\theta}{2}p_2\;\;\;,\;\;\;\hat{x}_{2 } = x_2+\frac{\theta}{2}p_1\;\;\;,\;\;\;\hat{x}_{3 } = x_3\;\;\;,\;\;\;\hat{p}_{i } = p_i
\;,\label{class}
\end{eqnarray}
and then, the Hamiltonian (\ref{ham}) takes the form
\begin{eqnarray}
{H}_\theta\left({\bar{x}},{\bar{p}}\right) &=& \frac{{{\bar{p}}^2}}{2m} - \frac{GmM}{\sqrt{\left(x_1-\frac{\theta}{2}p_2\right)^2 + 
\left(x_2+\frac{\theta}{2}p_1\right)^2 + x_3^2}} - \frac{m\Lambda c^2}{3}{\bar{r}}^2 \;+\nonumber\\
&-& \frac{m\Lambda c^2}{12}\theta^2\left(p_1^2+p_2^2\right) - \frac{m\Lambda c^2}{3}\theta L_3
\;; \label{hamosc}\\
L_3 &=& x_2p_1 - x_1p_2\;,\nonumber
\end{eqnarray}
while the corresponding equations of motion are given by
\begin{equation}
\left\{\begin{array}{rcl}
\dot{p}_1 &=& \frac{2}{3}m\Lambda c^2 x_1 - \frac{1}{3}{\Lambda c^2 m}\theta p_2 - \frac{GmM\left[x_1-\frac{\theta}{2}p_2\right]}{{\left[\left(x_1-\frac{\theta}{2}p_2\right)^2 + 
\left(x_2+\frac{\theta}{2}p_1\right)^2 + x_3^2\right]^{\frac{3}{2}}}} \;\\[5pt]
&~~&~\cr
\dot{p}_2 &=& \frac{2}{3}m\Lambda c^ x_2 + \frac{1}{3}{\Lambda c^2 m}\theta p_1 - \frac{GmM\left[x_2+\frac{\theta}{2}p_1\right]}{{\left[\left(x_1-\frac{\theta}{2}p_2\right)^2 + 
\left(x_2+\frac{\theta}{2}p_1\right)^2 + x_3^2\right]^{\frac{3}{2}}}} \;\\[5pt]
&~~&~\\[5pt]
\dot{p}_3 &=& \frac{2}{3}m\Lambda c^2 x_1 - \frac{GmMx_3}{{\left[\left(x_1-\frac{\theta}{2}p_2\right)^2 + 
\left(x_2+\frac{\theta}{2}p_1\right)^2 + x_3^2\right]^{\frac{3}{2}}}}  \;\\[5pt]
&~~&~\cr
\dot{x}_1 &=& \left[\frac{1}{m} - \frac{m\Lambda c^2}{6}\theta^2\right]p_1 +  \frac{GmM\theta\left[x_2+\frac{\theta}{2}p_1\right]}{2{\left[\left(x_1-\frac{\theta}{2}p_2\right)^2 + 
\left(x_2+\frac{\theta}{2}p_1\right)^2 + x_3^2\right]^{\frac{3}{2}}}}
\;\\[5pt]
&~~&~\cr
\dot{x}_2 &=& \left[\frac{1}{m} - \frac{m\Lambda c^2}{6}\theta^2\right]p_2 - \frac{GmM\theta\left[x_1-\frac{\theta}{2}p_2\right]}{2{\left[\left(x_1-\frac{\theta}{2}p_2\right)^2 + 
\left(x_2+\frac{\theta}{2}p_1\right)^2 + x_3^2\right]^{\frac{3}{2}}}}\;\\[5pt]
&~~&~\cr
\dot{x}_3 &=& \frac{p_3}{m} \;.
\label{equations}\end{array}\right.
\end{equation}
Obviously, for deformation parameter $\theta$ running to zero the all above formulas become commutative.

\subsection{First Lie-algebraic twist deformations}

In the case of the first twist deformation we put in formula (\ref{ssstar}) the indexes $\gamma$, $l$ and $k$ equal to 1, 2 and 3 respectively; then, we have
\begin{eqnarray}
\{\;\hat{x}_{1 },\hat{x}_{2 }\;\} = \frac{1}{\kappa_1}\hat{x}_3\;\;\;,\;\;\;\{\;\hat{x}_{1 },\hat{x}_{3 }\;\} = -\frac{1}{\kappa_1}\hat{x}_2\;\;\;,\;\;\;\{\;\hat{x}_{2},\hat{x}_{3 }\;\} = 0
\;. \label{newspace}
\end{eqnarray}
Further, we observe that the relations (\ref{newspace}) can be raised to the whole phase space in the following way \cite{miao}\footnote{Here we consider the so-called first type of deformed phase space for noncommutative space-time (\ref{newspace}) proposed in article \cite{miao}. It has been obtained as a solution of the proper condition for Jacobi identity.}
\begin{eqnarray}
&&\{\;\hat{x}_{1 },\hat{x}_{2 }\;\} = \hat{x}_3/\kappa_1\;\;\;,\;\;\;\{\;\hat{x}_{1 },\hat{x}_{3 }\;\} = -\hat{x}_2/\kappa_1\;\;\;,\;\;\;\{\;\hat{x}_{2},\hat{x}_{3 }\;\} = 0\; \label{p1}\\
&&\{\;\hat{x}_{1 },\hat{p}_{2 }\;\} = \hat{p}_3/\kappa_1\;\;\;,\;\;\;\{\;\hat{p}_{i},\hat{p}_{j }\;\} = 0\;\;\;,\;\;\;\{\;\hat{x}_{1 },\hat{p}_{3 }\;\} = -\hat{p}_2/\kappa_1\;, \label{phasespace1}\\
&&\{\;\hat{x}_{i },\hat{p}_{i }\;\} = 1\;\;\;,\;\;\;\{\;\hat{x}_{2 },\hat{p}_{3 }\;\} = 0 = \{\;\hat{x}_{3 },\hat{p}_{2 }\;\} = 
\{\;\hat{x}_{2 },\hat{p}_{1 }\;\} =
\{\;\hat{x}_{3 },\hat{p}_{1 }\;\}\;,\label{p2}
\end{eqnarray}
as well as we notice that the quantum variables $({\hat x}_i,{\hat p}_i)$ are realized by commutative ones $({x}_i,{p}_i)$ as \cite{meljanac}
\begin{eqnarray}
\hat{x}_{1 } = x_1-\frac{1}{\kappa_1}x_3p_2+\frac{1}{\kappa_1}x_2p_3\;\;\;,\;\;\;\hat{x}_{2 } = x_2\;\;\;,\;\;\;\hat{x}_{3 } = x_3\;\;\;,\;\;\;\hat{p}_{i } = p_i
\;. \label{classen}
\end{eqnarray}
Consequently, the Hamiltonian of the system takes the form 
\begin{eqnarray}
{H}_\theta\left({\bar{x}},{\bar{p}}\right) &=& \frac{1}{2m}{{\bar{p}}^2} - \frac{GmM}{\sqrt{\left(x_1-\frac{1}{\kappa_1}x_3p_2+\frac{1}{\kappa_1}x_2p_3\right)^2 + x_2^2 + x_3^2}} - \frac{m\Lambda c^2}{3}{\bar{r}}^2 \;+\nonumber\\
&-&  \frac{2m\Lambda c^2}{3\kappa_1}x_1L_1 - \frac{m\Lambda c^2}{\kappa_1^2}L_1^2
\;; \label{hamosclie1}\\
L_1 &=& x_2p_3-x_3p_2\;,\nonumber
\end{eqnarray}
while the canonical equations of motion are given by
\begin{equation}
\left\{\begin{array}{rcl}
\dot{p}_1 &=& \frac{2}{3}m\Lambda c^2 x_1 + \frac{2}{3\kappa_1}m\Lambda c^2 L_1 - \frac{GmM\left[x_1-\frac{1}{\kappa_1}x_3p_2+\frac{1}{\kappa_1}x_2p_3\right]}{\kappa_1{\left[\left(x_1-\frac{1}{\kappa_1}x_3p_2+\frac{1}{\kappa_1}x_2p_3\right)^2 + 
x_2^2 + x_3^2\right]^{\frac{3}{2}}}}  \;\\[5pt]
&~~&~\cr
\dot{p}_2 &=& \frac{2}{3}m\Lambda c^2 x_2 + \frac{2}{3\kappa_1}m\Lambda c^2x_1p_3 + \frac{2}{\kappa_1^2}m\Lambda c^2 L_1p_3 \;+ \;\\[5pt]
&~~&~\cr
&~~&~~~~~~~~~~~~~~~~~~~~~~~~~~~~~-\frac{GmM\left[\left[x_1-\frac{1}{\kappa_1}x_3p_2+\frac{1}{\kappa_1}x_2p_3\right]\frac{1}{\kappa_1}p_3+x_2\right]}{{\left[\left(x_1-\frac{1}{\kappa_1}x_3p_2+\frac{1}{\kappa_1}x_2p_3\right)^2 + 
x_2^2 + x_3^2\right]^{\frac{3}{2}}}} \;\\[5pt]
&~~&~\cr
\dot{p}_3 &=& \frac{2}{3}m\Lambda c^2 x_3 - \frac{2}{3\kappa_1}m\Lambda c^2 x_1p_2 - \frac{2}{\kappa_1^2}m\Lambda c^2 L_1p_2 \;+ \;\\[5pt]
&~~&~\cr
&~~&~~~~~~~~~~~~~~~~~~~~~~~~~~~~~-\frac{GmM\left[-\left[x_1-\frac{1}{\kappa_1}x_3p_2+\frac{1}{\kappa_1}x_2p_3\right]\frac{1}{\kappa_1}p_2+x_3\right]}{{\left[\left(x_1-\frac{1}{\kappa_1}x_3p_2+\frac{1}{\kappa_1}x_2p_3\right)^2 + 
x_2^2 + x_3^2\right]^{\frac{3}{2}}}}
\;\\[5pt]
&~~&~\cr
\dot{x}_1 &=& \frac{p_1}{m}\;\\[5pt]
&~~&~\cr
\dot{x}_2 &=& \frac{p_2}{m}  + \frac{2}{3\kappa_1}m\Lambda c^2 x_1x_3 + \frac{2}{\kappa_1^2}{m\Lambda c^2}
L_1x_3  -  \frac{GmMx_3\left[x_1-\frac{1}{\kappa_1}x_3p_2+\frac{1}{\kappa_1}x_2p_3\right]}{\kappa_1{\left[\left(x_1-\frac{1}{\kappa_1}x_3p_2+\frac{1}{\kappa_1}x_2p_3\right)^2 + 
x_2^2 + x_3^2\right]^{\frac{3}{2}}}} \;\\[5pt]
&~~&~\cr
\dot{x}_3 &=& \frac{p_3}{m}- \frac{2}{3\kappa_1}m\Lambda c^2 x_1x_2 - \frac{2}{\kappa_1^2}{m\Lambda c^2}
L_1x_2 +  \frac{GmMx_2\left[x_1-\frac{1}{\kappa_1}x_3p_2+\frac{1}{\kappa_1}x_2p_3\right]}{\kappa_1{\left[\left(x_1-\frac{1}{\kappa_1}x_3p_2+\frac{1}{\kappa_1}x_2p_3\right)^2 + 
x_2^2 + x_3^2\right]^{\frac{3}{2}}}}  \;.
\label{equationslie1}\end{array}\right.
\end{equation}
Of course, for deformation parameter $\kappa_1$ running to infinity the above model become commutative.

\subsection{Second Lie-algebraic twist deformation}

For the second Lie-algebraic twist deformation (see formula (\ref{ysesstar})) we take 
\begin{eqnarray}
\{\;\hat{x}_{1 },\hat{x}_{2 }\;\} = \frac{t}{\kappa_2}\;\;\;,\;\;\;\{\;\hat{x}_{1 },\hat{x}_{3 }\;\} =  0 = \{\;\hat{x}_{2 },\hat{x}_{3}\;\} \;, \label{newspacelie2}
\end{eqnarray}
and \cite{miao}\footnote{Here we consider the phase space for relations (\ref{newspacelie2}) proposed in \cite{miao} which satisfy the Jacobi identity.}
\begin{eqnarray}
\{\;\hat{x}_{i },\hat{x}_{j }\;\} = \theta_{ij}\;\;\;,\;\;\;\{\;\hat{x}_{i },\hat{p}_{j}\;\} = \delta_{ij}\;\;\;,\;\;\;\{\;\hat{p}_{i},\hat{x}_{j}\;\} = 0\;,\label{phasespacelie2}
\end{eqnarray}
where
\begin{eqnarray}
\hat{x}_{1 } = x_1-\frac{t}{2\kappa_2}p_2\;\;\;,\;\;\;\hat{x}_{2 } = x_2+\frac{t}{2\kappa_2}p_1\;\;\;,\;\;\;\hat{x}_{3 } = x_3\;\;\;,\;\;\;\hat{p}_{i } = p_i
\;.\label{classennn}
\end{eqnarray}
Then, the corresponding Hamiltonian function as well as the proper canonical equations look as follows 
\begin{eqnarray}
{H}_{\kappa_2}\left({\bar{x}},{\bar{p}},t\right) &=& \frac{{{\bar{p}}^2}}{2m} - \frac{GmM}{\sqrt{\left(x_1-\frac{t}{2\kappa_2}p_2\right)^2 + 
\left(x_2+\frac{t}{2\kappa_2}p_1\right)^2 + x_3^2}} - \frac{m\Lambda c^2}{3}{\bar{r}}^2 \;+\nonumber\\
&-& \frac{mt^2\Lambda c^2}{12\kappa_2^2}\left(p_1^2+p_2^2\right) - \frac{mt\Lambda c^2}{3\kappa_2}L_3
\;; \label{hamosclie2}\\
L_3 &=& x_2p_1 - x_1p_2\;,\nonumber
\end{eqnarray}
and
\begin{equation}
\left\{\begin{array}{rcl}
\dot{p}_1 &=& \frac{2}{3}m\Lambda c^2 x_1 - \frac{1}{3\kappa_2} {\Lambda c^2mt} p_2 - \frac{GmM\left[x_1-\frac{t}{2\kappa_2}p_2\right]}{{\left[\left(x_1-\frac{t}{2\kappa_2}p_2\right)^2 + 
\left(x_2+\frac{t}{2\kappa_2}p_1\right)^2 + x_3^2\right]^{\frac{3}{2}}}} \;\\[5pt]
&~~&~\cr
\dot{p}_2 &=& \frac{2}{3}m\Lambda c^2 x_2 + \frac{1}{3\kappa_2} {\Lambda c^2 mt}p_1 - \frac{GmM\left[x_2+\frac{t}{2\kappa_2}p_1\right]}{{\left[\left(x_1-\frac{t}{2\kappa_2}p_2\right)^2 + 
\left(x_2+\frac{t}{2\kappa_2}p_1\right)^2 + x_3^2\right]^{\frac{3}{2}}}} \;\\[5pt]
&~~&~\cr
\dot{p}_3 &=& \frac{2}{3}m\Lambda c^2 x_1 - \frac{GmMx_3}{{\left[\left(x_1-\frac{t}{2\kappa_2}p_2\right)^2 + 
\left(x_2+\frac{t}{2\kappa_2}p_1\right)^2 + x_3^2\right]^{\frac{3}{2}}}}  \;\\[5pt]
&~~&~\cr
\dot{x}_1 &=& \left[\frac{1}{m} - \frac{m\Lambda c^2 t^2}{6\kappa_2^2}\right]p_1 + \frac{GmMt\left[x_2+\frac{t}{2\kappa_2}p_1\right]}
{2\kappa_2{\left[\left(x_1-\frac{t}{2\kappa_2}p_2\right)^2 + 
\left(x_2+\frac{t}{2\kappa_2}p_1\right)^2 + x_3^2\right]^{\frac{3}{2}}}} \;\\[5pt]
&~~&~\cr
\dot{x}_2 &=& \left[\frac{1}{m} - \frac{m\Lambda c^2 t^2}{6\kappa_2^2}\right]p_2 - \frac{GmMt\left[x_1-\frac{t}{2\kappa_2}p_2\right]}
{2\kappa_2{\left[\left(x_1-\frac{t}{2\kappa_2}p_2\right)^2 + 
\left(x_2+\frac{t}{2\kappa_2}p_1\right)^2 + x_3^2\right]^{\frac{3}{2}}}}\;\\[5pt]
&~~&~\cr
\dot{x}_3 &=& \frac{p_3}{m} \;,
\label{equationslie2}\end{array}\right.
\end{equation}
respectively. Obviously, for deformation parameter $\kappa_2$ approaching infinity the above model become classical.

\section{Relations between models}

Let us now compare the provided in this paper models (\ref{equations}), (\ref{equationslie1}) and (\ref{equationslie2}). First of all, one should notice that the equations  (\ref{equationslie1}) are contrary to the remaining two systems highly nonlinear in $(x_i,p_i)$-variables. Besides, the Hamiltonian function for third model (\ref{hamosclie2}) is not conserved in time. Nevertheless, the all twisted Schwarzschild-(Anti-)de Sitter systems can be directly linked with use of so-called active control procedure \cite{conect1}, \cite{conectn}. In it's framework one can provide the proper dynamical coupling of the differently deformed particles such that for large times of the evolution their phase space trajectories becomes identical, i.e., the systems become synchronized (connected)\footnote{From the physical point of view, the above mentioned synchronization procedure gives an answer on the question: How should interact two cosmological particles moving in the presence of different (deformed) dynamics in order to their trajectories for large times become the same?}. Formally, such an interaction  
is described by the control functions which in the case of synchronization of the canonically deformed model with the second one look as follows\footnote{For details of finding the control functions see \cite{conect1}, \cite{conectn}.},\footnote{The trajectories $(x,p)$ and $(y,\pi)$ correspond to the master canonically (master) and Lie-algebraically (slave) deformed systems respectively while $K_1 = y_2\pi_3-y_3\pi_2$.},\footnote{The controllers (\ref{control1}) (the interaction terms) are added to the equations of motion (\ref{equationslie2}) and due to the Lyapunov theorem \cite{theorem} the systems are synchronized.}
\begin{equation}
\left\{\begin{array}{rcl}
u_{\theta,\kappa_1,\pi_1} &=&   \frac{2}{3}m\Lambda c^2 x_1 - \frac{1}{3}{\Lambda c^2 m}\theta p_2 - \frac{GmM\left[x_1-\frac{\theta}{2}p_2\right]}{{\left[\left(x_1-\frac{\theta}{2}p_2\right)^2 + 
\left(x_2+\frac{\theta}{2}p_1\right)^2 + x_3^2\right]^{\frac{3}{2}}}} + p_1 - \pi_1 \;+\;\\[5pt]
&~~&~\cr
&-&\frac{2}{3}m\Lambda c^2 y_1 - \frac{2}{3\kappa_1}m\Lambda c^2 K_1 + \frac{GmM\left[y_1-\frac{1}{\kappa_1}y_3\pi_2+\frac{1}{\kappa_1}y_2\pi_3\right]}{\kappa_1{\left[\left(y_1-\frac{1}{\kappa_1}y_3\pi_2+\frac{1}{\kappa_1}y_2\pi_3\right)^2 + 
y_2^2 + y_3^2\right]^{\frac{3}{2}}}}  \;\\[5pt]
&~~&~\cr
u_{\theta,\kappa_1,\pi_2} &=&  \frac{2}{3}m\Lambda c^2 x_2 + \frac{1}{3}{\Lambda c^2 m}\theta p_1 - \frac{GmM\left[x_2+\frac{\theta}{2}p_1\right]}{{\left[\left(x_1-\frac{\theta}{2}p_2\right)^2 + 
\left(x_2+\frac{\theta}{2}p_1\right)^2 + x_3^2\right]^{\frac{3}{2}}}} + p_2 - \pi_2\;+\;\\[5pt]
&~~&~\cr
&-&\frac{2}{3}m\Lambda c^2 y_2 - \frac{2}{3\kappa_1}m\Lambda c^2 y_1\pi_3 - \frac{2}{\kappa_1^2}m\Lambda c^2 K_1\pi_3 \;+ \;\\[5pt]
&~~&~\cr
&~~&~~~~~~~~~~~~~~~~~~~~~~~~~~~~~+\frac{GmM\left[\left[y_1-\frac{1}{\kappa_1}y_3\pi_2+\frac{1}{\kappa_1}y_2\pi_3\right]\frac{1}{\kappa_1}\pi_3+y_2\right]}{{\left[\left(y_1-\frac{1}{\kappa_1}y_3\pi_2+\frac{1}{\kappa_1}y_2\pi_3\right)^2 + 
y_2^2 + y_3^2\right]^{\frac{3}{2}}}} \;\\[5pt]
&~~&~\cr
u_{\theta,\kappa_1,\pi_3} &=& \frac{2}{3}m\Lambda c^2 x_1 - \frac{GmMx_3}{{\left[\left(x_1-\frac{\theta}{2}p_2\right)^2 + 
\left(x_2+\frac{\theta}{2}p_1\right)^2 + x_3^2\right]^{\frac{3}{2}}}} + p_3 - \pi_3\;+\;\\[5pt]
&~~&~\cr
&-&\frac{2}{3}m\Lambda c^2 y_3 + \frac{2}{3\kappa_1}m\Lambda c^2 y_1\pi_2 + \frac{2}{\kappa_1^2}m\Lambda c^2 K_1\pi_2 \;+ \;\\[5pt]
&~~&~\cr
&~~&~~~~~~~~~~~~~~~~~~~~~~~~~~~~~+\frac{GmM\left[-\left[y_1-\frac{1}{\kappa_1}y_3\pi_2+\frac{1}{\kappa_1}y_2\pi_3\right]\frac{1}{\kappa_1}\pi_2+y_3\right]}{{\left[\left(x_1-\frac{1}{\kappa_1}y_3\pi_2+\frac{1}{\kappa_1}y_2\pi_3\right)^2 + 
y_2^2 + y_3^2\right]^{\frac{3}{2}}}}
\label{control1}\end{array}\right.
\end{equation}
\begin{equation}
\left\{\begin{array}{rcl}
&~~&~\cr
u_{\theta,\kappa_1,y_1} &=& \left[\frac{1}{m} - \frac{m\Lambda c^2}{6}\theta^2\right]p_1 +  \frac{GmM\theta\left[x_2+\frac{\theta}{2}p_1\right]}{2{\left[\left(x_1-\frac{\theta}{2}p_2\right)^2 + 
\left(x_2+\frac{\theta}{2}p_1\right)^2 + x_3^2\right]^{\frac{3}{2}}}} + x_1 - y_1 - \frac{\pi_1}{m}\;\\[5pt]
&~~&~\cr
u_{\theta,\kappa_1,y_2} &=& \left[\frac{1}{m} - \frac{m\Lambda c^2}{6}\theta^2\right]p_2 - \frac{GmM\theta\left[x_1-\frac{\theta}{2}p_2\right]}{2{\left[\left(x_1-\frac{\theta}{2}p_2\right)^2 + 
\left(x_2+\frac{\theta}{2}p_1\right)^2 + x_3^2\right]^{\frac{3}{2}}}} + x_2 - y_2\;+\;\\[5pt]
&~~&~\cr
&-&\frac{\pi_2}{m} - \frac{2}{3\kappa_1}m\Lambda c^2 y_1y_3 - \frac{2}{\kappa_1^2}{m\Lambda c^2}
K_1y_3  + \frac{GmMy_3\left[y_1-\frac{1}{\kappa_1}y_3\pi_2+\frac{1}{\kappa_1}x_2p_3\right]}{\kappa_1{\left[\left(x_1-\frac{1}{\kappa_1}x_3p_2+\frac{1}{\kappa_1}x_2p_3\right)^2 + 
x_2^2 + x_3^2\right]^{\frac{3}{2}}}} \;\\[5pt]
&~~&~\cr
u_{\theta,\kappa_1,y_3} &=& \frac{p_3}{m} + x_3 - y_3 -\frac{\pi_3}{m} + \frac{2}{3\kappa_1}m\Lambda c^2 y_1y_2 + \frac{2}{\kappa_1^2}{m\Lambda c^2}
K_1y_2 \;+\;\\[5pt]
&~~&~\cr
&~~&~~~~~~~~~~~~~~~~~~~~~~~~~~~~~  - \frac{GmMy_2\left[y_1-\frac{1}{\kappa_1}y_3\pi_2+\frac{1}{\kappa_1}y_2\pi_3\right]}{\kappa_1{\left[\left(y_1-\frac{1}{\kappa_1}y_3\pi_2+\frac{1}{\kappa_1}y_2\pi_3\right)^2 + 
y_2^2 + y_3^2\right]^{\frac{3}{2}}}}\;.
\nonumber\end{array}\right.
\end{equation}
Besides, the active controllers $(u_{\theta,\kappa_2,\pi_i},u_{\theta,\kappa_2,y_i})$ 
coupling the first and third system as well as the functions $(u_{\kappa_1,\kappa_2,\pi_i},u_{\kappa_1,\kappa_2,y_i})$ combining the Lie-algebraically noncommutative models (\ref{equationslie1}) and (\ref{equationslie2}) can be find as well. However, due to the complicated form their presentation has been  
omitted in this paper. 

\section{Final remarks}

In this article we provide three twist-deformed Newtonian Schwarzschild-(Anti-)de Sitter models. They are defined on the Lie-algebraically as well as on the canonically noncommutative space-times respectively. Particularly, we find the corresponding Hamiltonian functions and the proper equations of motion. The synchronization of the models are discussed as well. 

It should be noted that the presented systems are quite interesting. They formally describe for example the impact of two different distance scales such as transplanckian (noncommutativity) and cosmological ($\Lambda$) scale on the dynamics of nonrelativistic particle moving in the central gravitational field. However, the better understanding of such a property of the models requires more investigations which are in progress.

\section*{Acknowledgments}

The author would like to thank J. Lukierski for valuable discussions.


\begin{thebibliography}{99}
\bibitem{kottler} F. Kottler, Ann. Phys. 361, 401 (1918)
\bibitem{bending} W. Rindler and M. Ishak, Phys. Rev. D 76, 043006 (2007); arXiv: 0709.2948 [astro-ph]
\bibitem{tests} H. Miraghaei, M. Nouri-Zonoz, Gen. Rel. Grav. 42, 2947 (2010); arXiv: 0810.2006 [gr-qc]
\bibitem{prefield} P. Kosinski, J. Lukierski, P. Maslanka, Phys. Rev. D {62}, 025004 (2000); arXiv: hep-th/9902037
\bibitem{qq13} M.R. Douglas and N.A. Nekrasov, Rev. Mod. Phys. {73}, 977 (2001); hep-th/0106048
\bibitem{qq14} R.J. Szabo, Phys. Rept. {378}, 207 (2003); hep-th/0109162
\bibitem{qq15} J.C. Wallet, J. Phys. Conf. Ser. {103}, 012007 (2008); arXiv: 0708.2471 [hep-th]
\bibitem{qq18} H. Grosse and R. Wulkenhaar, JHEP {0312}, 019 (2003); hep-th/0307017
\bibitem{qq20} A. Gere, T. Juric and J.C. Wallet, JHEP {1512}, 045 (2015); arXiv: 1507.08086 [hep-th]
\bibitem{field} M. Chaichian, P. Pre\v{s}najder and  A. Tureanu,
Phys. Rev. Lett. 94, 151602 (2005)
\bibitem{def1} M. Daszkiewicz, Acta Phys. Polon. B {47}, 2387 (2016); arXiv: 1610.08361 [physics.class-ph]
\bibitem{def1a} M. Daszkiewicz, Mod. Phys. Lett. A {32}, 1750075 (2017); arXiv: 1702.08702 [hep-th]
\bibitem{def1b} M. Daszkiewicz, Mod. Phys. Lett. A {33}, 1850100 (2018); arXiv: 1806.08357 [nlin.CD]
\bibitem{defn} M. Daszkiewicz {\it ''Noncommutative Sprott systems, their jerk dynamics, and their chaos synchronization by active control''} Talk given on The 32nd International Colloquium on Group Theoretical Methods in Physics (Group32), Prague 2018
\bibitem{mech} A. Deriglazov, JHEP 0303, 021 (2003); hep-th/0211105
\bibitem{qq4} B. Ivetic, S. Mignemi, A. Samsarov, Phys. Rev. A 93, 032109 (2016); arXiv: 1510.06196 [q-ph]
\bibitem{qq5} S. Mignemi, R. Strajn, \textit{"Quantum mechanics on a curved Snyder space"}; arXiv: 1501.01447 [hep-th]
\bibitem{qq8} T. Juric, S. Meljanac, A. Samsarov, J. Phys. Conf. Ser. 670, 012027 (2016); arXiv:1511.05592 [hep-th]
\bibitem{qq10} E. Harikumar, T. Juric, S. Meljanac, Phys. Rev. D 86, 045002 (2012); arXiv: 1203.1564 [hep-th]
\bibitem{qq12} D. Meljanac, S. Meljanac, D. Pikutic, K.S. Gupta, Phys. Rev. D 96, 105008 (2017); arXiv: 1703.09511 [hep-th]
\bibitem{qmq} M. Chaichian, M.M. Sheikh-Jabbari, A. Tureanu, Phys.
Rev. Lett. 86, 2716 (2001); hep-th/0010175
\bibitem{osc1} A. Kijanka, P. Kosinski, Phys. Rev. D {70}, 127702 (2004); arXiv: hep-th/0407246
\bibitem{osc2} M. Daszkiewicz, C.J. Walczyk, Acta Phys. Polon. B {40}, 293 (2009); arXiv: 0812.1264 [hep-th]
\bibitem{qm} M. Daszkiewicz, C.J. Walczyk, Phys. Rev. D {77}, 105008 (2008); arXiv: 0802.3575 [math-ph]
\bibitem{class1}S. Zakrzewski, \textit{"Poisson Structures on the Poincare
group"}; q-alg/9602001
\bibitem{class2}
Y. Brihaye, E. Kowalczyk, P. Maslanka, \textit{"Poisson-Lie structure on Galilei
group"}; math/0006167
\bibitem{nnh}M. Daszkiewicz, Mod. Phys. Lett. A27 (2012) 1250083; arXiv: 1205.0319 [hep-th]
\bibitem{oeckl}R. Oeckl, J. Math. Phys. 40, 3588 (1999)
\bibitem{chi}M. Chaichian, P.P. Kulish, K. Nashijima, A. Tureanu, Phys. Lett. B
604, 98 (2004); hep-th/0408069
\bibitem{dasz1}M. Daszkiewicz,
Mod. Phys. Lett. A 23, 505 (2008); arXiv: 0801.1206 [hep-th]
\bibitem{kappaP}J. Lukierski, A. Nowicki, H. Ruegg and V.N. Tolstoy, Phys. Lett.
B 264, 331 (1991)
\bibitem{kappaG}S. Giller, P. Kosinski, M. Majewski, P. Maslanka
and J. Kunz, Phys. Lett. B 286, 57 (1992)
\bibitem{lie1}J. Lukierski and M. Woronowicz, Phys. Lett. B 633, 116 (2006); hep-th/0508083
\bibitem{qdef}O. Ogievetsky, W.B.  Schmidke, J. Wess, B. Zumino, Comm. Math. Phys.
150, 495 (1992)
\bibitem{paolo}
P. Aschieri, L. Castellani, A.M. Scarfone, Eur. Phys. J. C 7, 159
(1999); q-alg/9709032
\bibitem{drin} V.G. Drinfeld, Soviet Math. Dokl. 32, 254 (1985); Algebra i
Analiz (in Russian), 1, Fasc. 6, p. 114 (1989)
\bibitem{qg} S. Majid {\it ''Foundations of quantum group theory''} Cambridge University Press, April 2000
\bibitem{bad2} A. Larranaga, \textit{''Geodesic Structure of the Noncommutative Schwarzschild Anti-de Sitter Black Hole I: Timelike Geodesics''}; arXiv: 1110.0778 [gr-qc]
\bibitem{badn} J.M. Romero, J.A. Santiago, Mod. Phys. Lett. A 20, 781 (2005); arXiv: hep-th/0310266
\bibitem{conect1} M.C. Ho, Y.C. Hung, Phys. Lett. A {301}, 424 (2002)
\bibitem{conectn} H.K. Chen, Chaos, Solitons and Fractals {23}, 1245 (2005)
\bibitem{planck}G. Amelino-Camelia, Living Rev. Rel. {16}, 5 (2013); arXiv: 0806.0339 [gr-qc]
\bibitem{Qgrav}C. Kiefer, ''Quantum Gravity'', Oxford University Press, 2007
\bibitem{QGC}D. Blas, O. Pujolas, S. Sibiryakov, JHEP 1104:018 (2011); arXiv: 1007.3503 [hep-th]
\bibitem{bopp}F. Bopp, Ann. Inst. H. Poincare, 1581 (1956)
\bibitem{miao}Y.G. Miao, X.D. Wang, S.J. Yu, Annals Phys. 326, 2091 (2011); arXiv: 0911.5227 [math-ph]
\bibitem{meljanac}T. Juric, S. Meljanac, D. Pikutic,  	Eur. Phys. J. C 75, 528 (2015); arXiv: 1506.04955 [hep-th]
\bibitem{theorem}A.M.  Lyapunov,  \textit{”The  General  Problem  of  the  Stability  of  Motion”}  (In  Russian),Doctoral  dissertation,  Univ.  Kharkov  1892.  English  translation:  \textit{”Stability  of  Motion”}, Academic Press, New-York and London, 1966
\end{thebibliography}
\end{document}